\pdfoutput=1
\documentclass[aps,arxiv,twocolumn,superscriptaddress]{revtex4-1}
\usepackage{graphicx}
\usepackage{float}
\usepackage{amsmath}
\usepackage{amssymb}
\usepackage{mathtools}
\usepackage[utf8]{inputenc}
\usepackage[T1]{fontenc}
\usepackage[version=3]{mhchem}
\usepackage{microtype}

\newlength{\colwidth}
\begin{document}

\title{The electronic-structure origin of cation disorder in
  transition-metal oxides}

\author{Alexander Urban}
\email{aurban@berkeley.edu}
\affiliation{%
  Department of Materials Science and Engineering,
  University of California, Berkeley, CA, USA}
\author{Aziz Abdellahi}
\thanks{A.~Abdellahi, S.~Dacek, and N.~Artrith contributed equally to
  this work.}
\affiliation{%
  Department of Materials Science and Engineering,
  Massachusetts Institute of Technology, Cambridge, MA, USA.}
\author{Stephen Dacek}
\thanks{A.~Abdellahi, S.~Dacek, and N.~Artrith contributed equally to
  this work.}
\affiliation{%
  Department of Materials Science and Engineering,
  Massachusetts Institute of Technology, Cambridge, MA, USA.}
\author{Nongnuch Artrith}
\thanks{A.~Abdellahi, S.~Dacek, and N.~Artrith contributed equally to
  this work.}
\affiliation{%
  Department of Materials Science and Engineering,
  University of California, Berkeley, CA, USA}
\author{Gerbrand Ceder}
\email{gceder@berkeley.edu}
\affiliation{%
  Department of Materials Science and Engineering,
  University of California, Berkeley, CA, USA}
\affiliation{%
  Materials Science Division, Lawrence Berkeley National
  Laboratory, Berkeley, CA, USA}
\date{\today}

\begin{abstract}
  Cation disorder is an important design criterion for technologically
  relevant transition-metal (TM) oxides, such as radiation-tolerant
  ceramics and Li-ion battery electrodes.
  In this letter, we use a combination of first-principles calculations,
  normal mode analysis, and band-structure arguments to pinpoint a
  specific electronic-structure effect that influences the stability of
  disordered phases.
  We find that the electronic configuration of a TM ion determines to
  which extent the structural energy is affected by site distortions.
  This mechanism explains the stability of disordered phases with large
  ionic radius differences and provides a concrete guideline for the
  discovery of novel disordered compositions.
\end{abstract}

\maketitle

Substitutional disorder is a common phenomenon in transition-metal (TM)
oxides and is known to affect structural and electronic properties.
For example, cation disorder induces structural amorphization in
\ce{La2Zr2O7} pyrochlores~\cite{prl94-2005-025505}, controls the
magnetoresistance in Fe-Mo perovskites~\cite{prl86-2001-2443}, and
affects the critical temperature of \ce{La2CuO4}
superconductors~\cite{n394-1998-157}.
In rocksalt-type Li-TM oxides, cation disorder determines the Li-ion
conductivity~\cite{aem-2014-1400478, sci343-2014-519}, an important
performance measure for Li-ion battery cathodes.

The technological relevance of cation-disordered oxides creates the
desire to predict whether a given composition is likely to be
disordered.
While high-throughput first-principles computations are useful to screen
specific composition spaces for stable disordered
compounds~\cite{prb79-2009-104203, nc6-2015-8485, aem-2016-1600488,
  cm28-2016-6484}, a better understanding of the origin of cation
disorder might lead to simple design criteria so that time-consuming
computations can be avoided.

For metallic alloys, the Hume-Rothery rules predict that species with
similar electronegativity form a solid solution when their atomic radii
differ no more than 15\%~\cite{zkcm91-1935-23, pearson2013}, but this
simple heuristic does not directly translate to covalent and ionic
materials such as oxides.
For example, cation disorder in pyrochlores has been extensively
studied~\cite{s289-2000-748}, and while the ionic radii are an important
factor for the tendency to disorder~\cite{cjc46-1968-859},
species-dependent differences in the metal-oxygen
bonding~\cite{jacs83-2000-1873} and electronic-structure
effects~\cite{prb79-2009-104203} have prevented the formulation of
heuristic rules to reliably predict disorder.
Likewise, B-site cation disorder in \ce{$A$($B'B''$)O3} perovskites has
been linked to the similarity of the ionic radius and charge of the
\ce{$B'$} and \ce{$B''$} species, but these two parameters alone cannot
explain all experimentally observed trends~\cite{jacs74-1991-2846,
  armr38-2008-369}.
On the other hand, cation-disordered Li-TM oxides with large ionic
radius and charge differences are known, e.g.,
\ce{LiNi_{0.5}Ti_{0.5}O2}~\cite{jps185-2008-534} where the Shannon radii
of \ce{Li^+}, \ce{Ni^2+}, and \ce{Ti^4+} are~76,~69, and 61~pm,
respectively~\cite{aca32-1976-751}, which seems to contradict the
present understanding of the origin of cation disorder.

In an ordered structure, a TM has a single or few local high-symmetry
environments, whereas a disordered structure has a large number of
distinct low-symmetry environments.
Hence, the ability of a TM to disorder will to some extent depend on how
it can accommodate such a variety of
environments~\cite{jacs83-2000-1873}.
We demonstrate in this letter that such an adaptability is determined by
the TM's electronic structure.
We show that $d^{0}$ TMs promote disorder while other
$d$-electron configurations, especially the $d^{6}$ configuration,
strongly prohibit disorder.
This mechanism explains the formation of solid solutions with cation
species that exhibit considerable ionic radius differences.


\begin{figure*}
  \centering
  \includegraphics[width=2\colwidth]{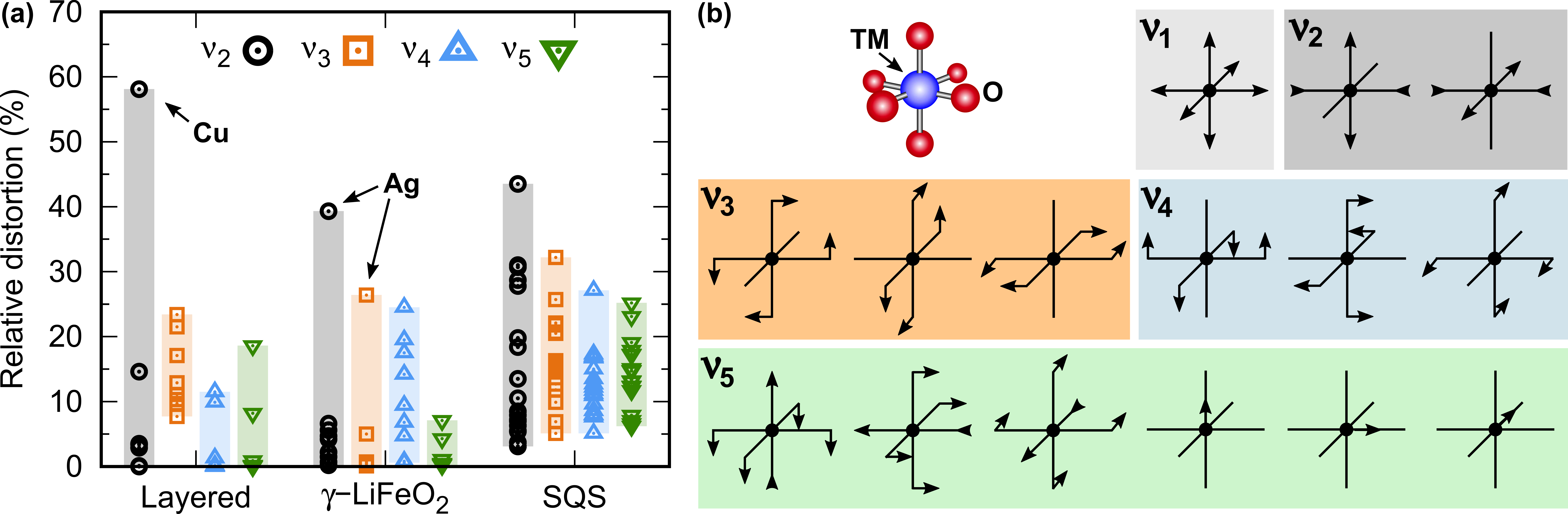}
  \caption{\label{fig:modes-SQS-vs-others}%
    \textbf{(a)}~TM site distortion in the \ce{LiTMO2} ground state
    structures (layered \ce{\alpha-NaFeO2} structure and
    \ce{\gamma-LiFeO2} structure) and in the special quasi-random
    structures (SQS) of all first and second-row TMs except Mn and Tc.
    The distortions are decomposed into contributions from different
    normal mode symmetry groups.
    The first normal mode ($\nu_{1}$) corresponds to an isotropic
    scaling and is not considered.
    $\nu_{2}$ is the Jahn-Teller distortion, $\nu_{3}$ corresponds
    to bending distortions, $\nu_{4}$ to twisting, and $\nu_{5}$
    describes the displacement of the TM from the center of the
    site.
    \textbf{(b)}~Schematic of the normal modes of an octahedral TM site
    grouped by symmetry (rotations and translations are not shown).}
\end{figure*}
As a case study we focus on the Li/TM disorder in cation-disordered
\ce{LiTMO2} compounds, which have recently attracted interest as Li-ion
battery cathode materials~\cite{sci343-2014-519, ec60-2015-70,
  pnsa112-2015-7650, CM-2015-Li-Ti-Fe-O, cm28-2016-416, cc52-2016-2051}.
The presence of Li cations with no valence electrons as one of the
components simplifies the analysis by focusing on the electronic
configuration of the TM.

For all first and second row TMs the ground state \ce{LiTMO2} structure
is either the layered \ce{\alpha-NaFeO2} structure or the
\ce{\gamma-LiFeO2} structure with the exception of \ce{LiMnO2} which
forms an orthorhombic structure~\cite{jpcs48-1987-97, aem-2016-1600488}.
The cation sites in these rocksalt-type Li-TM oxides are octahedral.
For a large number of TMs in the \ce{LiTMO2} composition we calculate
the energy and relaxed atomic configuration for the ground state and the
disordered structure as represented by a Special Quasi-Random Structure
(SQS)~\cite{prl65-1990-353, prb76-2007-144204}.


Structures and energies were obtained from spin-polarized
density-functional theory (DFT) calculations~\cite{pr136-64-846,
  pr140-65-A1133, jcp140-2014-18A301} using the PBE
functional~\cite{prl80-1998-891, prl78-97-1396}, PAW
pseudopotentials~\cite{prB50-1994-17953} as implemented in
VASP~\cite{prB54-1996-11169, cms6-1996-15}, and k-point meshes with a
density of 1000 divided by the number of atoms~\cite{cms68-2013-314}.
A Hubbard-U correction~\cite{prB44-1991-943, prB52-1995-R5467,
  prB57-1998-1505, cms50-2011-2295} was employed to correct the DFT
self-interaction error (see Table~\ref{tbl:U-values} in the appendix).
All DFT energies and atomic forces were converged to 0.05~meV per atom
and 50~meV\,\AA{}$^{-1}$, respectively, and the plane-wave cutoff was
520~eV.


To more directly understand the response of the TM electronic states on
distortions of the local atomic environment we represent the
displacement of the TM and oxygen atoms from their ideal positions in
terms of the 21 normal coordinates of the octahedral \ce{MO6} structure
given in Table~\ref{tbl:Oh-normal-coordinates}.
For the octahedral point group, the normal coordinates (excluding
rotations and translations) belong to the five different groups
$\nu_{1}$ through $\nu_{5}$ shown in
\textbf{Fig.~\ref{fig:modes-SQS-vs-others}b}.
The symmetric stretching mode $\nu_{1}$ corresponds to an isotropic
scaling and does not contribute to any distortion.
The asymmetric stretching modes of type $\nu_{2}$ are the modes of the
Jahn--Teller (JT) distortion~\cite{prsl161-1937-220, prsl164-1938-117,
  jcp7-1939-72}.
The modes of types $\nu_{3}$ and $\nu_{4}$ describe \emph{bending} and
\emph{twisting}, respectively, and the modes in $\nu_{5}$ describe the
\emph{displacement} of the cation from the center of the site.

The amplitude of the four symmetry-breaking normal modes around the
metal cations in the two ordered ground state structures and disordered
SQS are shown in \textbf{Fig.~\ref{fig:modes-SQS-vs-others}a} for all
first- and second-row TMs except Mn and Tc.
The relative distortions are given by the coefficients of the normal
coordinates in the representation of the distorted octahedron $D$ based
on an ideal octahedron $O$, $D = O + \sum_{i}c_{i} \widetilde{Q}_{i}$,
where $\widetilde{Q}_{i}$ are the normalized normal coordinates of
Table~\ref{tbl:Oh-normal-coordinates}.
For each group of normal coordinates, the largest coefficient $c_{i}$ is
plotted in \textbf{Fig.~\ref{fig:modes-SQS-vs-others}a} and listed in
Table~\ref{tbl:TM-site-distortions}.
As seen in \textbf{Fig.~\ref{fig:modes-SQS-vs-others}a}, the site
distortions in most of the ordered layered and \ce{\gamma-LiFeO2}
structures range from 0~to~25\% with the exception of \ce{LiCuO2}
(layered) and \ce{LiAgO2} (\ce{\gamma-LiFeO2} structure) which exhibit
strong site distortions of type $\nu_{2}$ because of the preference of
$d^{8}$ \ce{Cu^3+} and \ce{Ag^3+} for square planar coordination.
Further, the other ordered compounds exhibit no or only minor
distortions (<\,5\%) in the JT~mode ($\nu_{2}$) and the TM-displacement
mode~($\nu_{5}$).
Contributions of the TM-displacement mode ($\nu_{5}$) are only
significant for the second-row TMs Nb, Mo, Ru, and Ag.

The situation is different for the SQS, in which all four types of
distortions are present for all of the TMs.
In addition, the magnitude of the distortions is on average greater, and
the contribution of the JT mode is greater than 5\% for most of the TMs.
Interestingly, the amplitude of the TM-displacement mode $\nu_{5}$ is
also between 5\% and 25\% in each SQS.
Hence, even cations that reside in nearly undistorted sites in their
ground-state structure, e.g., ($d^{6}$) \ce{Co^3+} and \ce{Rh^3+}, will
be subject to site distortions in a cation-disordered structure.

\begin{figure}
  \centering
  \includegraphics[width=\colwidth]{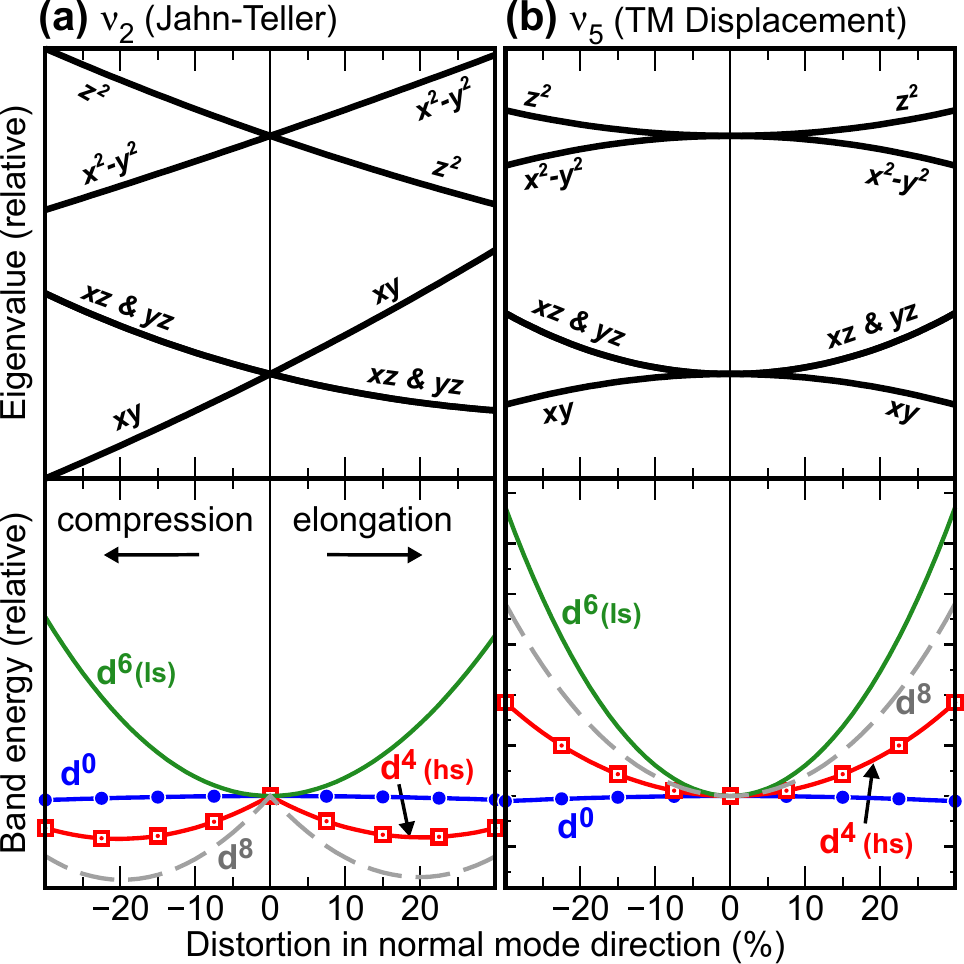}
  \caption{\label{fig:TB}%
    Change of the electronic states (top) and the band energy (bottom)
    upon distortion of an octahedral TM site in the direction of
    \textbf{(a)}~the Jahn--Teller mode ($\nu_{2}$) and \textbf{(b)}~the
    TM-displacement mode ($\nu_{5}$).
    The band energies for four electronic configurations are shown:
    $d^{0}$ (blue circles), $d^{4}$ high spin (hs, red squares), $d^{6}$
    low spin (ls, green line), and $d^{8}$ (gray dashed line).
    The energies shown in panel~\textbf{(a)} are based on the
    conventional Jahn--Teller mode (compression/elongation in $z$
    direction) which is a linear combination of the normal coordinates
    of type~$\nu_{2}$ shown in Fig.~\ref{fig:modes-SQS-vs-others}b.
    The energy and distortion scales are equal for both normal modes.
    The labels in the top panels indicate the TM $d$~orbitals that
    contribute most for distortions along the Cartesian $z$ direction.}
\end{figure}
To interpret the effect of these site distortions on the energy of the
disordered phase we consider the band-sum expression of the total
energy~\cite{prB39-1989-12520, prB31-85-1770, Foulkes1987},
$
  U
  = E_{\textup{band}}
  + \mathcal{D}
  \quad\text{with}\quad
  E_{\textup{band}} = \sum_{i}^{\textup{occ.}} \varepsilon_{i}
$,
where $\varepsilon_{i}$ are the eigenvalues of the Kohn--Sham
single-electron Hamiltonian (i.e., the energies of the electronic
eigenstates), and the sum over the eigenvalues of all occupied
electronic states is the \emph{band energy} $E_{\textup{band}}$.
The term $\mathcal{D}$ contains a \emph{double-counting} correction,
\emph{electrostatics}, and contributions from the
\emph{exchange--correlation} functional, and it is mostly determined by
pairwise interactions~\cite{prB39-1989-12520}.
$E_{\textup{band}}$ captures the distance and angle-dependent change of
the electronic states with the local structural
environment~\cite{prB39-1989-12520}.
Hence, to understand the energy trends in disordered structures we ought
to analyze the response of $E_{\textup{band}}$ on site distortions.

We seek a qualitative picture of band-energy trends for general TM
oxides rather than quantitative energies for select compositions (as
these can be obtained directly from DFT), so that a simple tight-binding
(TB) model of the electronic structure is most appropriate.
As such, we construct a model Hamiltonian for an octahedral TM site
based on the oxygen $p$ and TM $d$ hydrogen-like atomic orbitals in the
spirit of the extended Hückel method~\cite{jcp39-1963-1397},
$
  H_{ij} = K\,\frac{H_{ii} + H_{jj}}{2}\,S_{ij}
$,
where $S_{ij}\!=\!\!\int\phi_{i}(r)\phi_{j}(r)\,\mathrm{d}r$ is the
overlap of orbitals $\phi_{i}$ and $\phi_{j}$, $K=1.75$ is the
Wolfsberg--Helmholtz constant~\cite{jcp20-1952-837, jcp39-1963-1397},
and the diagonal Hamilton matrix elements $H_{ii}$ are the ionization
potentials (IP) of the oxygen $p$ and TM $d$ valence states.
The angular dependence of the overlap integrals is obtained from the
tables by Slater and Koster~\cite{pr94-1954-1498} for reference
integrals with appropriate radial symmetry (see section~\ref{sec:TB} in
the appendix).

Note that the actual choice of the (TM-species dependent) IPs only
affects the absolute energy of the electronic states $\varepsilon_{i}$
but not the relative change with different distortions, so for the
present discussion we only require that the oxygen $p$~level lies lower
in energy than the TM $d$~level.
Thus, the choice made for the IPs does not limit the generality of the
trends discussed in the following.

\begin{figure*}
  \centering
  \includegraphics[width=2\colwidth]{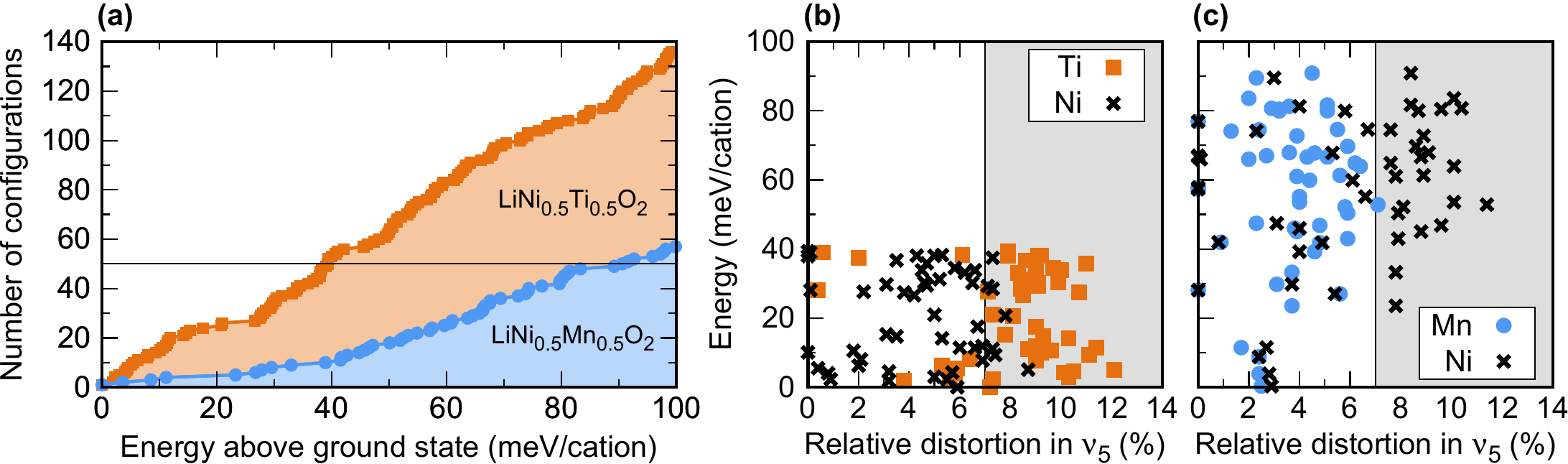}
  \caption{\label{fig:distortion-NiTi-vs-NiMn}%
    \textbf{(a)}~Number of \ce{LiNi_{0.5}Ti_{0.5}O2} and
    \ce{LiNi_{0.5}Mn_{0.5}O2} configurations within a 100~meV/cation
    from the ground state out of 469~distinct
    configurations with up to four formula units.
    \textbf{(b)}~Energy per cation and TM site distortion of the 50~most
    stable \textbf{(b)}~\ce{LiNi_{0.5}Ti_{0.5}O2} and
    \textbf{(c)}~\ce{LiNi_{0.5}Mn_{0.5}O2} configurations.
    Only TM-displacement ($\nu_{5}$) distortions are considered. }
\end{figure*}

\textbf{Figure~\ref{fig:TB}} shows the change of the electronic states
and of $E_{\textup{band}}$ that results from distortions in the
directions of the JT mode ($\nu_{2}$) and the TM-displacement mode
($\nu_{5}$) predicted by the TB model.
The impact of bending ($\nu_{3}$) and twisting ($\nu_{4}$) distortions
on $E_{\textup{band}}$ (Fig.~\ref{fig:E_band_TB}) is small compared to
the JT ($\nu_{2}$) and displacement ($\nu_{5}$) modes.
Unlike the bending and twisting modes, the JT~distortion and the TM
displacement directly affect the TM-O bond length, explaining the
stronger effect of these two modes on the energy levels.

Only the five electronic states corresponding to the TM $d$ orbitals are
shown in \textbf{Fig.~\ref{fig:TB}}, i.e., the two $e_{g}$ and three
$t_{2g}$ states in the ideal octahedral crystal field, as the dependence
of the lower-lying states on the site distortions is negligible in
comparison.
The band energies for four $d$-electron configurations, $d^{0}$, $d^{4}$
(high spin), $d^{6}$ (low spin), and $d^{8}$ are shown in
\textbf{Fig.~\ref{fig:TB}}, and the remaining $d$-electron
configurations can be found in Fig.~\ref{fig:E_band_TB}.
Note that $E_{\textup{band}}$ also contains contributions from the lower
lying states.

Note that, while the covalent character of the TM-O bond varies with the
TM species, the formal TM valence state in oxides usually corresponds to
the correct $d$-electron count~\cite{esl5-2002-A145}.


As seen in \textbf{Fig.~\ref{fig:TB}}, distortions of the TM site in
either JT or TM-displacement mode lower the symmetry such that the
degenerate $e_{g}$ and $t_{2g}$ levels split.
In the case of the JT mode, the energy of the electronic states changes
approximately linearly, whereas TM displacement only gives rise to
quadratic and higher-order changes.
As a consequence, for TM~displacements with amplitudes $<$5\%, the
relative change of the electronic states is small compared to the effect
of a JT distortion with a comparable amplitude.
However, for distortions with amplitudes of $\geq$10\%, the magnitude of
the energy change resulting from both normal modes is comparable.

The same general trend is seen in the band energies
(\textbf{Fig.~\ref{fig:TB}}), as the JT distortion results in a linear
change of $E_{\textup{band}}$ for some $d$-electron configurations while
the effect of the TM~displacement is at most quadratic.
The net energy change for TM~displacements with amplitudes $>$10\% is,
nevertheless, larger than for JT~distortions for most $d$-electron
counts.
JT~distortions can increase or reduce $E_{\textup{band}}$ depending on
the electronic configuration of the cation.
TM~displacements result in a steep increase of $E_{\textup{band}}$ for
TMs with more than four valence $d$ electrons and slightly stabilize
$d^{1}$ and $d^{2}$ TMs (Fig.~\ref{fig:E_band_TB}).
Note that distortions in either JT or TM-displacement mode result in a
strong increase of $E_{\textup{band}}$ for $d^{6}$ (low spin) and
$d^{10}$ configurations.
Finally, in the absence of $t_{2g}$ and $e_{g}$ electrons, i.e., for the
$d^{0}$ configuration, $E_{\textup{band}}$ solely depends on the
lower-lying oxygen-dominated orbitals that are always occupied.
As a result, $d^{0}$ TMs are least sensitive with respect to TM site
distortions, and the variation of $E_{\textup{band}}$ is only minor.

A key conclusion of the original paper by Jahn and
Teller~\cite{prsl161-1937-220} is that, to first order, only distortions
of type $\nu_{2}$ can affect the electronic energy.
Our results are fully consistent with the JT theorem, as the
TM-displacement mode ($\nu_{5}$) only brings about at most second order
changes of $E_{\textup{band}}$.
When the TM ion is displaced from the center of its octahedral site, one
TM-O bond is elongated and a second TM-O bond is compressed, and only
the difference of both effects is seen in $E_{\textup{band}}$.
%

Most importantly, here we find that the second-order contributions to
the band energy by the TM~displacement are not negligible for the large
distortions that occur in cation-disordered Li-TM oxides with amplitudes
between 5~and~25\% (\textbf{Fig.~\ref{fig:modes-SQS-vs-others}}).
Additionally, from the band energies in \textbf{Fig.~\ref{fig:TB}} it is
obvious that not only JT~active TMs are affected by site distortions, as
significant energy contributions occur from distortions for any TM ion
with more than zero $d$ electrons.
Since only the band energy of $d^{0}$ TMs is insensitive with respect to
distortions, we conclude that $d^{0}$ TMs tolerate disordered phases
even for relatively large ionic radius differences.

As seen in Fig.~\ref{fig:E_band_DFT}, the DFT band energies of actual
$d^{0}$, $d^{6}$, and $d^{8}$ Li-TM oxides (\ce{LiYO2}, \ce{LiRhO2}, and
\ce{LiAgO2}) follow precisely the trend predicted by the TB model for
TM-displacement distortions.

\begin{table}[htb]
  \centering
  \caption{\label{tbl:DO-LiMO2}%
    Cation-disordered \ce{Li_{1+x}TM_{1-x}O2} made by conventional
    solid-state synthesis.  \ce{Li_{1.211}Mo_{0.467}Cr_{0.3}O_{2}},
    forms in the layered (\ce{\alpha-NaFeO2}) structure and converts to
    the disordered (NaCl) structure upon Li extraction and simultaneous
    oxidation of \ce{Mo^5+} to \ce{Mo^6+}~\cite{sci343-2014-519}.
    The other materials form directly in the disordered rocksalt
    structure.
    $d^{0}$ TM cations are highlighted in bold font.}\vspace{0.5em}
  \begin{tabular}{ll}
    \hline\hline
    \textbf{Composition}
    & \textbf{TM Cations} \\
    \hline
    \ce{LiMO2} (M\,=\,Ti, Fe)~\cite{jpcs48-1987-97}
      & \ce{Ti^3+}, \ce{Fe^3+} \\
    \ce{Li_{1+x}V_{2}O_{5}}~\cite{jps34-1991-113}
      & \ce{V^3+}, \textbf{\ce{V^5+}} \\
    \ce{LiM_{0.5}Ti_{0.5}O2} (M\,=\,Fe, Ni)~\cite{jes150-2003-A638, jps185-2008-534}
      & \ce{M^2+}, \textbf{\ce{Ti^4+}} \\
    \ce{Li_{1.211}Mo_{0.467}Cr_{0.3}O_{2}}~\cite{sci343-2014-519}\,$^{a}$
      & \ce{Mo^5+}, \ce{Cr^3+} \\
    \ce{Li_{1.25}Nb_{0.25}Mn_{0.5}O2}~\cite{ec60-2015-70}
      & \textbf{\ce{Nb^5+}}, \ce{Mn^3+} \\
    \ce{Li_{1.3}Nb_{0.3+x}M_{0.4-x}O2} (M\,=\,Mn, Fe, Co, Ni)~\cite{pnsa112-2015-7650}
      & \textbf{\ce{Nb^5+}}, \ce{M^3+} \\
    \ce{Li_{1+x}Ti_{2x}Fe_{1-3x}O_{2}}~\cite{CM-2015-Li-Ti-Fe-O}
      & \textbf{\ce{Ti^4+}}, \ce{Fe^3+} \\
    \ce{Li_{1.6-4x}Mo_{0.4-x}Ni_{5x}O2}~\cite{cm28-2016-416}
      & \textbf{\ce{Mo^6+}}, \ce{Ni^2+} \\
    \ce{Li_{1.3}Nb_{0.3}V_{0.4}O2}~\cite{cc52-2016-2051}
      & \textbf{\ce{Nb^5+}}, \ce{V^3+} \\
    \ce{LiCo_{0.5}Zr_{0.5}O2}~\cite{aem-2016-1600488}
      & \ce{Co^2+}, \textbf{\ce{Zr^4+}} \\
    \hline\hline
    \multicolumn{2}{l}{%
      $^{a}$\,Forms in layered structure but disorders upon Li extraction.}
  \end{tabular}
\end{table}
\textbf{Table~\ref{tbl:DO-LiMO2}} lists published cation-disordered
Li-TM oxides that were made by conventional solid-state
synthesis~\cite{jpcs48-1987-97, jps34-1991-113, jes150-2003-A638,
  jps185-2008-534, sci343-2014-519, ec60-2015-70, pnsa112-2015-7650,
  CM-2015-Li-Ti-Fe-O, cm28-2016-416, cc52-2016-2051, aem-2016-1600488}.
Indeed, most of the compositions contain one TM species in a formal
valence state corresponding to the $d^{0}$ electronic configuration:
\ce{Ti^4+}, \ce{V^5+}, \ce{Zr^4+}, \ce{Nb^5+}, or \ce{Mo^6+}.
The Li-Mo-Cr oxide of reference~\onlinecite{sci343-2014-519} forms in
the layered structure but becomes cation disordered when Li is extracted
and \ce{Mo^5+} is oxidized to $d^{0}$ \ce{Mo^6+}.
The only cation-disordered compositions that do not contain $d^{0}$ TM
species are stoichiometric \ce{LiTiO2} ($d^{1}$ \ce{Ti^3+}) and
\ce{LiFeO2} ($d^{5}$ \ce{Fe^3+})~\cite{jpcs48-1987-97}.
In the case of \ce{LiFeO2}, calorimetry measurements showed that the
disordered \ce{\alpha-LiFeO2} phase is in fact significantly higher
($\sim$90\,meV) in energy than the ordered \ce{\gamma-LiFeO2} ground
state~\cite{jssc178-2005-1230}, and the formation of the $\alpha$-phase
during synthesis has been attributed to kinetic
reasons~\cite{jssc178-2005-1230}.
A similar mechanism might be responsible for the stabilization of
cation-disordered \ce{LiTiO2}.
Hence, the literature supports our hypothesis that $d^{0}$ cations
promote cation disorder.

Finally, to understand how the presence of $d^{0}$ cations within a
composition with several TM species can stabilize disordered structures,
we systematically enumerated atomic configurations of
\ce{LiNi_{0.5}Mn_{0.5}O2}, which is known to form a layered ground-state
structure~\cite{cl30-2001-744}, and \ce{LiNi_{0.5}Ti_{0.5}O2}, which is
cation disordered at typical synthesis
temperatures~\cite{jps185-2008-534}.
For both compositions, the DFT energies of 469~symmetrically distinct
atomic configurations with up to 8~cation sites were computed.
\textbf{Figure~\ref{fig:distortion-NiTi-vs-NiMn}a} shows the number of
atomic configurations within an energy range of 100~meV/cation above the
ground state for both oxides.
As seen in the figure, the number of configurations within this energy
interval is far greater for \ce{LiNi_{0.5}Ti_{0.5}O2}
(136~configuration) than for \ce{LiNi_{0.5}Mn_{0.5}O2}
(57~configurations).
This means, for the Ti-containing composition more atomic orderings are
thermally accessible at synthesis conditions.
The origin of this very different energetic behavior becomes obvious
when the TM site distortions are considered:
\textbf{Figure~\ref{fig:distortion-NiTi-vs-NiMn}b~and~c} show the
TM-displacement distortions in the 50~most stable configurations of both
materials.
As seen in the figure, the $d^{0}$ \ce{Ti^4+} cations accommodate large
site distortions allowing the \ce{Ni^2+} sites to remain close to their
preferred geometry.
Since $d^{0}$ cations are less sensitive with respect to site
distortions, the energy of these 50 configurations is within
40~meV/cation from the computational ground state.
In contrast, the energy of the \ce{LiNi_{0.5}Mn_{0.5}O2} configurations
increases rapidly with the relative distortion of the $d^{8}$ \ce{Ni^2+}
and $d^{3}$ \ce{Mn^4+} sites.
Note that some distortion of the Ni and Mn sites is tolerable, as the
band energy increases quadratically with the amplitude of $\nu_{5}$
(\textbf{Fig.~\ref{fig:TB}b}).
Hence, not only do $d^{0}$ cations have a low energy penalty in
distorted sites, their flexibility to distort allows the other TM
cations to minimize their distortions.


In conclusion, we identified a specific electronic-structure effect
that is responsible for the stabilization of cation-disordered phases in
lithium-transition-metal oxides with large cation size differences.
We showed that the strong transition-metal site distortions that occur
when these compositions disorder give rise to significant second-order
energy contributions.
As a consequence, $d^{6}$ transition metals are strongly destabilized in
cation-disordered phases, whereas $d^{0}$ transition metals can tolerate
such distortions with very low energy cost.
Owing to this tolerance, $d^{0}$ species can absorb site distortions in
mixed compositions, even when the cation sizes differ significantly.
At the example of technologically relevant lithium transition-metal
oxides, we show that this insight can function as a concrete guideline
for the design of novel cation-disordered compositions.
While our numerical data focused on lithium-transition-metal oxides, we
believe that the specific mechanism presented here by which disorder
comes at lower energy cost when the TM can accommodate the distorted
site more easily, will be more generally applicable to other oxides with
octahedral cations.

\section*{Acknowledgments}

This work was supported by the Robert Bosch Corporation and by Umicore
Specialty Oxides and Chemicals and by the Assistant Secretary for Energy
Efficiency and Renewable Energy, Office of Vehicle Technologies of the
U.S.\ Department of Energy under contract no. DE-AC02–05CH11231,
subcontract no. 7056411.
This work used the Extreme Science and Engineering Discovery Environment
(XSEDE), which is supported by National Science Foundation grant number
ACI-1053575.
In addition, resources of the National Energy Research Scientific
Computing Center, a DOE Office of Science User Facility supported by the
Office of Science of the U.S. Department of Energy under Contract
No. DE-AC02-05CH11231, are gratefully acknowledged.


\bibliography{AlexUrban-bibliography}{}
\bibliographystyle{aipnum4-1}

\onecolumngrid
\appendix
\clearpage

\section{Bond integrals used in the tight-binding model}
\label{sec:TB}

\begin{figure}[H]
  \centering
  \includegraphics[width=\textwidth]{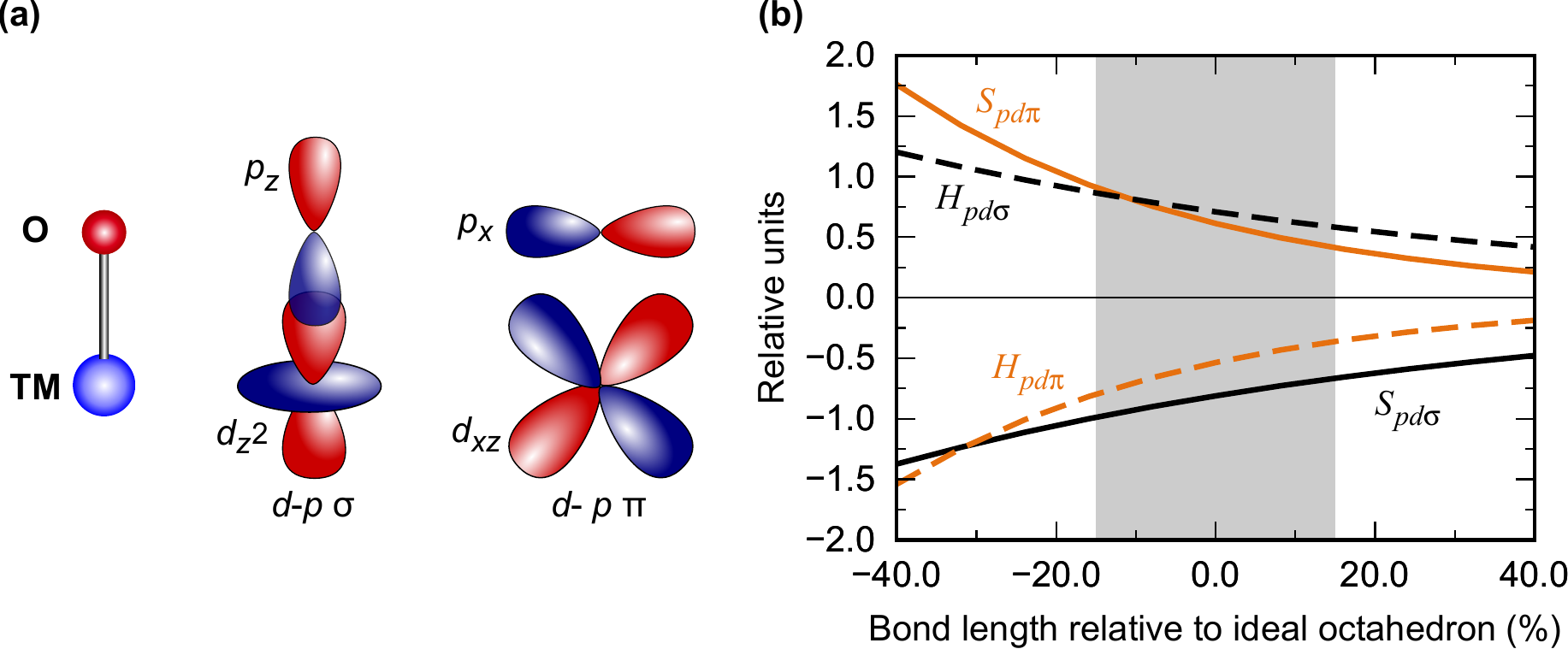}
  \caption{\label{fig:bonds}%
    \textbf{(a)}~Schematic of the reference bonds between the
    transition-metal $d$ and oxygen $p$ orbitals (Cartesian $z$
    direction).  All TM-O bonds for arbitrary geometries can be
    expressed in terms of these reference bonds following the approach
    by Slater and Koster~\cite{pr94-1954-1498}.  \textbf{(b)}~Plot of
    the reference overlap and bond integrals.  The gray shaded region
    indicates the TM-O bond lengths that actually occur in the
    considered structures.}
\end{figure}

Our tight-binding (TB) model assumes the usual hydrogen-like atomic
orbitals $\{\phi_{i}\}$, i.e., each atomic orbital is the product of a
radial function and a (real-valued) spherical harmonic function.
Only the bonds between the transition-metal $d$ orbitals and the oxygen
$p$ orbitals are considered.
Since the atomic-orbital radial functions decay exponentially with the
distance from the atomic center, the same is true for the overlap of two
orbitals $\phi_{i}$ and $\phi_{j}$
\begin{align}
  S_{ij}
  = \int\phi_{i}(r)\phi_{j}(r)\,\mathrm{d}r
  \label{eq:S_ij}
\end{align}
and we therefore chose the form of Slater-type
orbitals~\citep{pr36-1930-57} to describe the overlap integrals
$S_{pd\sigma}$ and $S_{pd\pi}$ of the two TM-O reference bonds
(\textbf{Fig.~\ref{fig:bonds}a})
\begin{align}
  S_{b}(r)
  = N e^{-\zeta_{b}\, r}
  \quad\text{with the normalization constant}\quad
  N = 2\, \zeta_{b} \sqrt{\zeta_{b}}
  \label{eq:STO}
  \quad ,
\end{align}
where $\zeta_{b}$ is a constant that controls the exponential decay.
The $\pi$ bond decays faster than the $\sigma$ bond, so that we chose
$\zeta_{pd\pi}=2\zeta_{pd\sigma}$ with $\zeta_{pd\sigma}=$\,1.32.
The overlap and bond integrals resulting from this choice are shown in
\textbf{Fig.~\ref{fig:bonds}b}.
We confirmed that all of our conclusions are robust with respect to the
choice of the $\zeta_{b}$ ratio and are not affected if the value of
$\zeta_{pd\pi}$ is varied between $1\zeta_{pd\sigma}$ and
$3\zeta_{pd\sigma}$.

\begin{table}[H]
  \centering
  \caption{\label{tbl:U-values}%
    Hubbard~U corrections used in DFT+U calculations.
    With the exception of the values for Mo, the U values are identical
    to those by Jain~et~al.~\cite{cms50-2011-2295}.
    The U values were fitted to measured binary formation enthalpies and
    are particularly well-suited for the calculation of phase
    stabilities.}%
  \vspace{0.5em}
  \begin{tabular}{lc|lc}
    \hline\hline
    \multicolumn{1}{c}{\textbf{TM}} &
    \multicolumn{1}{c|}{\textbf{U value (eV)}} &
    \multicolumn{1}{c}{\textbf{TM}} &
    \multicolumn{1}{c}{\textbf{U value (eV)}} \\
    \hline
      Ag & 1.50 &  Mn & 3.90 \\
      Co & 3.40 &  Mo & 4.38 \\
      Cr & 3.50 &  Nb & 1.50 \\
      Cu & 4.00 &  Ni & 6.00 \\
      Fe & 4.00 &  V  & 3.10 \\
    \hline\hline
  \end{tabular}
\end{table}

\newpage

\begin{figure}[H]
  \centering
  \includegraphics[width=0.3\textwidth]{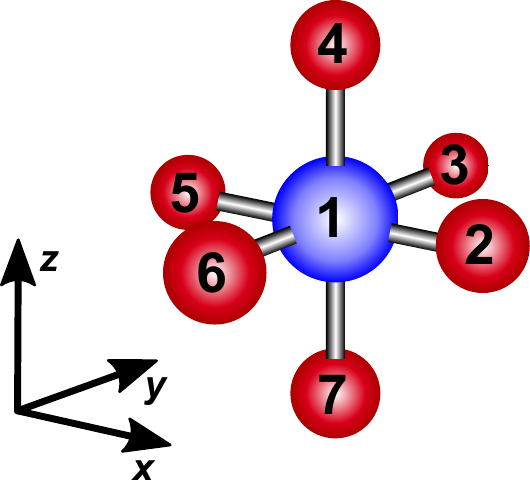}
  \caption{\label{fig:XY6}%
    Schematic showing the atom numbering convention used in
    Table~\ref{tbl:Oh-normal-coordinates}.  Red balls indicate oxygen
    atoms and the blue ball indicates the TM atom.}
\end{figure}

\begin{table}[H]
  \centering
  \caption{\label{tbl:Oh-normal-coordinates}%
    Normal coordinates of the octahedral \ce{MO6} site (not normalized or
    orthogonalized).
  }
  \vspace{0.5em}
  \scriptsize
\begin{tabular}{lrrr|rrr|rrr|rrr|rrr|rrr|rrr}
  \hline
  \hline
  &
  \multicolumn{3}{c}{\textbf{Atom~1}} &
  \multicolumn{3}{c}{\textbf{Atom~2}} &
  \multicolumn{3}{c}{\textbf{Atom~3}} &
  \multicolumn{3}{c}{\textbf{Atom~4}} &
  \multicolumn{3}{c}{\textbf{Atom~5}} &
  \multicolumn{3}{c}{\textbf{Atom~6}} &
  \multicolumn{3}{c}{\textbf{Atom~7}} \\
  \hline
  &
  \multicolumn{1}{c}{$\Delta{}x$} & \multicolumn{1}{c}{$\Delta{}y$} & \multicolumn{1}{c}{$\Delta{}z$} &
  \multicolumn{1}{c}{$\Delta{}x$} & \multicolumn{1}{c}{$\Delta{}y$} & \multicolumn{1}{c}{$\Delta{}z$} &
  \multicolumn{1}{c}{$\Delta{}x$} & \multicolumn{1}{c}{$\Delta{}y$} & \multicolumn{1}{c}{$\Delta{}z$} &
  \multicolumn{1}{c}{$\Delta{}x$} & \multicolumn{1}{c}{$\Delta{}y$} & \multicolumn{1}{c}{$\Delta{}z$} &
  \multicolumn{1}{c}{$\Delta{}x$} & \multicolumn{1}{c}{$\Delta{}y$} & \multicolumn{1}{c}{$\Delta{}z$} &
  \multicolumn{1}{c}{$\Delta{}x$} & \multicolumn{1}{c}{$\Delta{}y$} & \multicolumn{1}{c}{$\Delta{}z$} &
  \multicolumn{1}{c}{$\Delta{}x$} & \multicolumn{1}{c}{$\Delta{}y$} & \multicolumn{1}{c}{$\Delta{}z$} \\
\hline \multicolumn{22}{l}{\textbf{Symmetric stretching modes}} \\ \hline
$Q_{1}$  &  0 &  0 &  0 & +1 &  0 &  0 &  0 & +1 &  0 &  0 &  0 & +1 & -1 &  0 &  0 &  0 & -1 &  0 &  0 &  0 & -1 \\
\hline \multicolumn{22}{l}{\textbf{Assymmetric stretching modes}} \\ \hline
$Q_{2}$  &  0 &  0 &  0 & -1 &  0 &  0 &  0 &  0 &  0 &  0 &  0 & +1 & +1 &  0 &  0 &  0 &  0 &  0 &  0 &  0 & -1 \\
$Q_{3}$  &  0 &  0 &  0 & -1 &  0 &  0 &  0 & +1 &  0 &  0 &  0 &  0 & +1 &  0 &  0 &  0 & -1 &  0 &  0 &  0 &  0 \\
\hline \multicolumn{22}{l}{\textbf{Bending modes}} \\ \hline
$Q_{4}$  &  0 &  0 &  0 &  0 & +1 &  0 & +1 &  0 &  0 &  0 &  0 &  0 &  0 & -1 &  0 & -1 &  0 &  0 &  0 &  0 &  0 \\
$Q_{5}$  &  0 &  0 &  0 &  0 &  0 &  0 &  0 &  0 & -1 &  0 & -1 &  0 &  0 &  0 &  0 &  0 &  0 & +1 &  0 & +1 &  0 \\
$Q_{6}$  &  0 &  0 &  0 &  0 &  0 & -1 &  0 &  0 &  0 & -1 &  0 &  0 &  0 &  0 & +1 &  0 &  0 &  0 & +1 &  0 &  0 \\
\hline \multicolumn{22}{l}{\textbf{Twisting modes}} \\ \hline
$Q_{7}$  &  0 &  0 &  0 &  0 & -1 &  0 &  0 &  0 &  0 &  0 & +1 &  0 &  0 & -1 &  0 &  0 &  0 &  0 &  0 & +1 &  0 \\
$Q_{8}$  &  0 &  0 &  0 &  0 &  0 & -1 &  0 &  0 & +1 &  0 &  0 &  0 &  0 &  0 & -1 &  0 &  0 & +1 &  0 &  0 &  0 \\
$Q_{9}$  &  0 &  0 &  0 &  0 &  0 &  0 & +1 &  0 &  0 & -1 &  0 &  0 &  0 &  0 &  0 & +1 &  0 &  0 & -1 &  0 &  0 \\
\hline \multicolumn{22}{l}{\textbf{Displacement modes}} \\ \hline
$Q_{10}$ &  0 &  0 &  0 &  0 & +1 &  0 &  0 & -2 &  0 &  0 & +1 &  0 &  0 & +1 &  0 &  0 & -2 &  0 &  0 & +1 &  0 \\
$Q_{11}$ &  0 &  0 &  0 &  0 &  0 & +1 &  0 &  0 & +1 &  0 &  0 & -2 &  0 &  0 & +1 &  0 &  0 & +1 &  0 &  0 & -2 \\
$Q_{12}$ &  0 &  0 &  0 & -2 &  0 &  0 & +1 &  0 &  0 & +1 &  0 &  0 & -2 &  0 &  0 & +1 &  0 &  0 & +1 &  0 &  0 \\
$Q_{13}$ &  0 & +4 &  0 &  0 & -1 &  0 &  0 &  0 &  0 &  0 & -1 &  0 &  0 & -1 &  0 &  0 &  0 &  0 &  0 & -1 &  0 \\
$Q_{14}$ & +4 &  0 &  0 &  0 &  0 &  0 & -1 &  0 &  0 & -1 &  0 &  0 &  0 &  0 &  0 & -1 &  0 &  0 & -1 &  0 &  0 \\
$Q_{15}$ &  0 &  0 & -4 &  0 &  0 & +1 &  0 &  0 & +1 &  0 &  0 &  0 &  0 &  0 & +1 &  0 &  0 & +1 &  0 &  0 &  0 \\
\hline \multicolumn{22}{l}{\textbf{Translations}} \\ \hline
$Q_{16}$ & +1 &  0 &  0 & +1 &  0 &  0 & +1 &  0 &  0 & +1 &  0 &  0 & +1 &  0 &  0 & +1 &  0 &  0 & +1 &  0 &  0 \\
$Q_{17}$ &  0 & +1 &  0 &  0 & +1 &  0 &  0 & +1 &  0 &  0 & +1 &  0 &  0 & +1 &  0 &  0 & +1 &  0 &  0 & +1 &  0 \\
$Q_{18}$ &  0 &  0 & +1 &  0 &  0 & +1 &  0 &  0 & +1 &  0 &  0 & +1 &  0 &  0 & +1 &  0 &  0 & +1 &  0 &  0 & +1 \\
\hline \multicolumn{22}{l}{\textbf{Rotations}} \\ \hline
$Q_{19}$ &  0 &  0 &  0 &  0 &  0 &  0 &  0 &  0 & +1 &  0 & -1 &  0 &  0 &  0 &  0 &  0 &  0 & -1 &  0 & +1 &  0 \\
$Q_{20}$ &  0 &  0 &  0 &  0 &  0 & +1 &  0 &  0 &  0 & -1 &  0 &  0 &  0 &  0 & -1 &  0 &  0 &  0 & +1 &  0 &  0 \\
$Q_{21}$ &  0 &  0 &  0 &  0 & +1 &  0 & -1 &  0 &  0 &  0 &  0 &  0 &  0 & -1 &  0 & +1 &  0 &  0 &  0 &  0 &  0 \\
\hline
\hline
\end{tabular}
\end{table}

\newpage

\begin{table}[H]
  \centering
  \newcommand{\gLFO}{$\gamma$-LiFeO$_{2}$}
  \newcommand{\aNFO}{$\alpha$-NaFeO$_{2}$}
  \newcommand{\oLMO}{$o$-LiMnO$_{2}$}
  \caption{\label{tbl:TM-site-distortions}%
    Values of the TM site distortions visualized in Fig.~1 of the main
    manuscript.  The orthorhombic \ce{LiMnO2} structure is not shown in
    Fig.~1 but included here for completeness.
    The distortion values are the contributions of
    the different normal coordinates to the normal mode representation
    of the distorted TM sites, i.e., the expansion coefficients for each
    normalized normal coordinate of Table~\ref{tbl:Oh-normal-coordinates}.}
  \vspace{0.5em}
\begin{tabular}{llrrrr|crrrr}
  \hline
  \hline
  \multicolumn{1}{c}{\bfseries TM} &
  \multicolumn{1}{c}{\bfseries Structure} &
  \multicolumn{1}{c}{$\boldsymbol{\nu^{~}_{2}}$} &
  \multicolumn{1}{c}{$\boldsymbol{\nu^{~}_{3}}$} &
  \multicolumn{1}{c}{$\boldsymbol{\nu^{~}_{4}}$} &
  \multicolumn{1}{c}{$\boldsymbol{\nu^{~}_{5}}$} &
  \multicolumn{1}{c}{\bfseries Structure} &
  \multicolumn{1}{c}{$\boldsymbol{\nu^{~}_{2}}$} &
  \multicolumn{1}{c}{$\boldsymbol{\nu^{~}_{3}}$} &
  \multicolumn{1}{c}{$\boldsymbol{\nu^{~}_{4}}$} &
  \multicolumn{1}{c}{$\boldsymbol{\nu^{~}_{5}}$} \\
  \hline
Ag & \gLFO    & 0.393 & 0.264 & 0.094 & 0.071 & SQS & 0.307 & 0.207 & 0.167 & 0.151 \\
Cd & \gLFO    & 0.007 & 0.001 & 0.006 & 0.002 & SQS & 0.037 & 0.069 & 0.051 & 0.062 \\
Co & \aNFO    & 0.000 & 0.129 & 0.000 & 0.000 & SQS & 0.184 & 0.127 & 0.112 & 0.191 \\
Cr & \aNFO    & 0.001 & 0.097 & 0.013 & 0.009 & SQS & 0.063 & 0.163 & 0.079 & 0.123 \\
Cu & \aNFO    & 0.581 & 0.234 & 0.000 & 0.000 & SQS & 0.278 & 0.152 & 0.150 & 0.079 \\
Fe & \gLFO    & 0.002 & 0.005 & 0.175 & 0.003 & SQS & 0.079 & 0.118 & 0.080 & 0.124 \\
Mn & \oLMO    & 0.201 & 0.060 & 0.083 & 0.061 & SQS & 0.198 & 0.205 & 0.173 & 0.117 \\
Mo & \gLFO    & 0.015 & 0.050 & 0.068 & 0.043 & SQS & 0.054 & 0.161 & 0.109 & 0.117 \\
Nb & \aNFO    & 0.035 & 0.111 & 0.099 & 0.186 & SQS & 0.072 & 0.159 & 0.135 & 0.172 \\
Ni & \aNFO    & 0.146 & 0.103 & 0.002 & 0.002 & SQS & 0.135 & 0.155 & 0.117 & 0.066 \\
Rh & \aNFO    & 0.000 & 0.171 & 0.000 & 0.001 & SQS & 0.310 & 0.218 & 0.127 & 0.179 \\
Ru & \aNFO    & 0.034 & 0.215 & 0.115 & 0.082 & SQS & 0.435 & 0.322 & 0.170 & 0.231 \\
Sc & \gLFO    & 0.022 & 0.001 & 0.195 & 0.005 & SQS & 0.070 & 0.147 & 0.096 & 0.147 \\
Ti & \gLFO    & 0.042 & 0.002 & 0.008 & 0.004 & SQS & 0.054 & 0.099 & 0.086 & 0.132 \\
V  & \aNFO    & 0.029 & 0.077 & 0.001 & 0.001 & SQS & 0.085 & 0.148 & 0.109 & 0.148 \\
Y  & \gLFO    & 0.047 & 0.007 & 0.245 & 0.011 & SQS & 0.105 & 0.221 & 0.168 & 0.173 \\
Zn & \gLFO    & 0.066 & 0.001 & 0.142 & 0.002 & SQS & 0.031 & 0.051 & 0.076 & 0.071 \\
Zr & \gLFO    & 0.056 & 0.002 & 0.047 & 0.003 & SQS & 0.055 & 0.153 & 0.122 & 0.131 \\
  \hline
  \hline
\end{tabular}
\end{table}

\newpage

\begin{figure}[H]
  \centering
  \includegraphics[width=\textwidth]{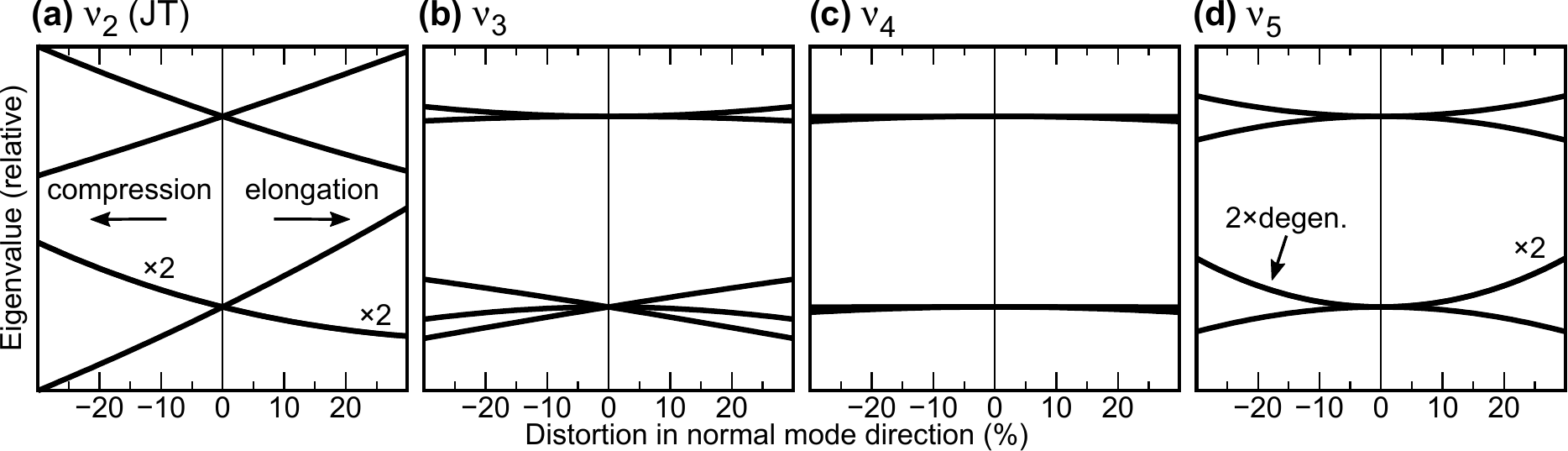}
  \caption{\label{fig:eval_TB}%
    Tight-binding eigenvalues for all symmetry-breaking normal modes.}
\end{figure}

\begin{figure}[H]
  \centering
  \includegraphics[width=0.66\textwidth]{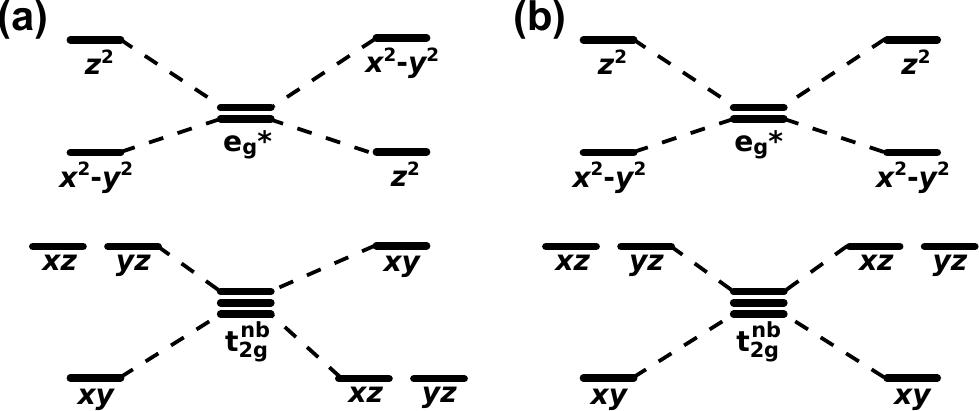}
  \caption{\label{fig:level-splitting}%
    Schematic of the molecular orbital level splitting on distortion in
    the \textbf{(a)}~Jahn-Teller mode ($\nu_{2}$) and
    \textbf{(b)}~transition-metal displacement mode ($\nu_{5}$).  The
    labels refer to the transition-metal $d$ orbital that contributes
    most to the state.  The vertical spacing does not reflect actual
    energy units.}
\end{figure}

\newpage

\begin{figure}[H]
  \centering
  \includegraphics[width=\textwidth]{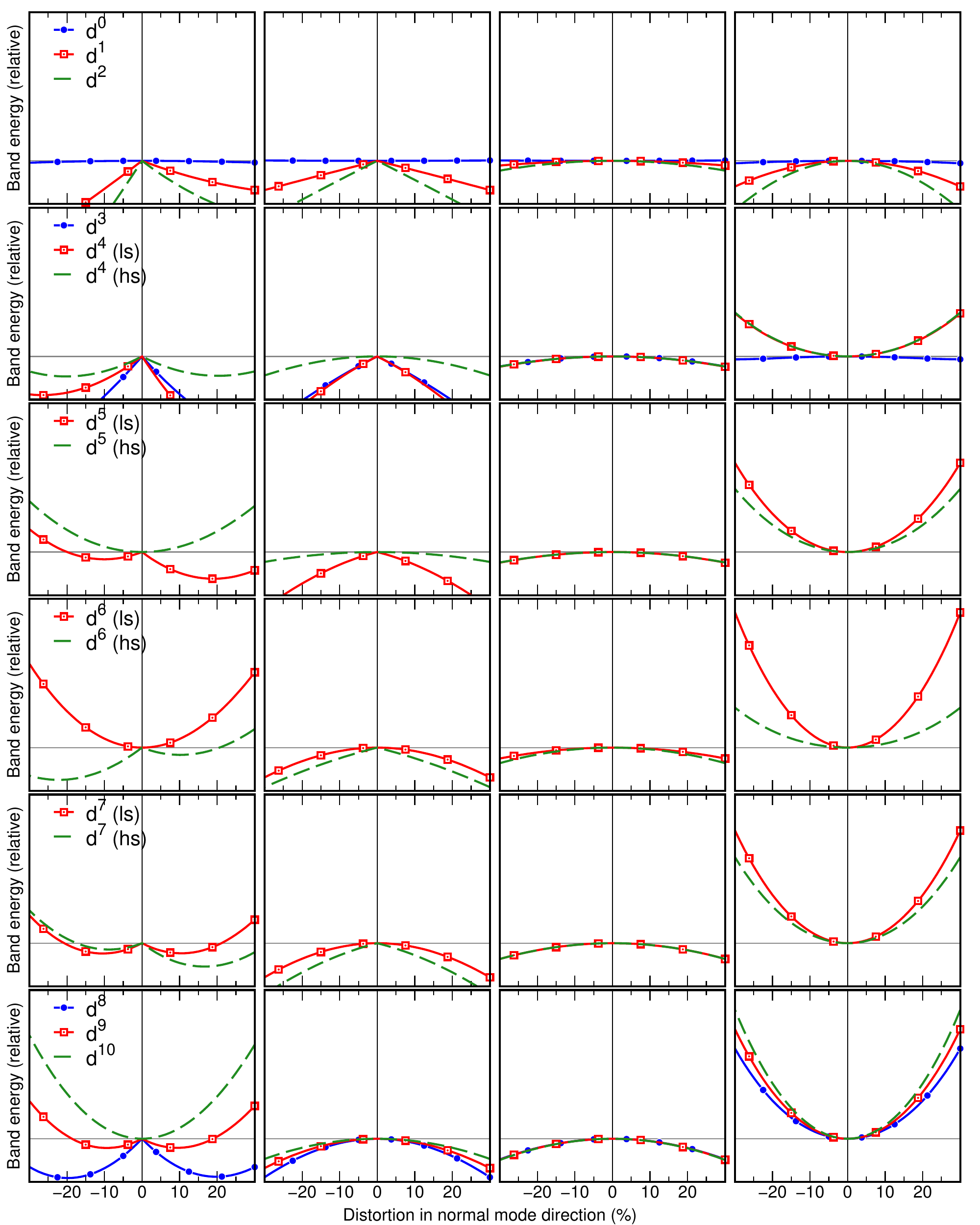}
  \caption{\label{fig:E_band_TB}%
    Tight-binding band energies for all $d$-electron configurations and
    spin states upon distortion in the direction of the five
    symmetry-breaking normal modes.}
\end{figure}

\begin{figure}[H]
  \centering
  \includegraphics[width=0.5\textwidth]{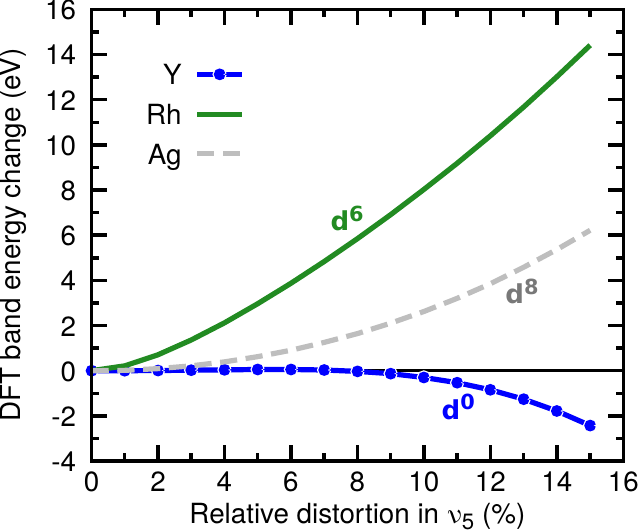}
  \caption{\label{fig:E_band_DFT}%
    Change of the density-functional theory (DFT) band energies of
    \ce{LiYO2}, \ce{LiRhO2}, and \ce{LiAgO2} upon distortion of the
    transition-metal site in the TM-displacement mode ($\nu_{5}$).
    To ensure a global energy reference, all DFT eigenvalues were
    calculated relative to the energy of the oxygen 2$s$ states which
    does not significantly hybridize with the valence
    bands of the 4$d$ TMs.
    The $d$-electron configuration of the transition metal cations is
    $d^{0}$ for \ce{Y^3+}, $d^{6}$ (low-spin) for \ce{Rh^3+}, and
    $d^{8}$ for \ce{Ag^3+}.
    The band energies shown in the figure are one concrete example of
    the general band energy trends shown in Fig.~2b of the main
    manuscript.}
\end{figure}

\end{document}